\documentclass[aps,pra,reprint,superscriptaddress,10pt,showpacs,a4paper]{revtex4-1}
\usepackage{amsmath,amssymb,amsfonts}
\usepackage{mathtools}
\usepackage{graphicx}
\usepackage{txfonts}
\usepackage{color, xcolor}

\usepackage[unicode,breaklinks]{hyperref}
\hypersetup{
    unicode=true,
    a4paper=true,
    plainpages=false,
    pdftitle={Response of a Superfluid to a Disordered Potential},
    pdfauthor={R. Doran, A. J. Groszek, and T. P. Billam},
    pdfsubject={Response of a Superfluid to a Disordered Potential},
    colorlinks=true,
    linkcolor=blue,
    citecolor=blue,
    filecolor=black,
    urlcolor=blue
}
\usepackage{tikz}

\newcommand{\secreft}[1]{{Section~\ref{#1}}}

\newcommand{\rr}{\mathbf{r}}
\newcommand{\vv}{\mathbf{v}}
\newcommand{\kk}{\mathbf{k}}

\newcommand{\pp}{\mathbf{p}}

\newcommand{\Arg}{\text{Arg}}
\newcommand{\vcrit}{v_{\mathrm{crit}}}
\newcommand{\vobst}{v_{\mathrm{obst}}}

\begin{document}
\title{Critical Velocity and Arrest of a Superfluid in a Point-Like Disordered Potential}

\author{R. Doran\address{r.doran@newcastle.ac.uk}}
\affiliation{Joint Quantum Centre (JQC) Durham--Newcastle, Department of Mathematics, Statistics and Physics, Newcastle University, Newcastle upon Tyne, NE1 7RU, UK}
\author{A. J. Groszek}
\affiliation{ARC Centre of Excellence in Future Low-Energy Electronics Technologies,
School of Mathematics and Physics, University of Queensland, St Lucia, QLD 4072, Australia}
\affiliation{ARC Centre of Excellence for Engineered Quantum Systems, School of Mathematics and Physics, University of Queensland, Saint Lucia QLD 4072, Australia}
\author{T. P. Billam}
\affiliation{Joint Quantum Centre (JQC) Durham--Newcastle, Department of Mathematics, Statistics and Physics, Newcastle University, Newcastle upon Tyne, NE1 7RU, UK}

\date{\today}

\begin{abstract}
  Superfluid flow past a potential barrier is a well studied problem in ultracold Bose gases, however, fewer studies have considered the case of flow through a disordered potential. Here we consider the case of a superfluid flowing through a channel containing multiple point-like barriers, randomly placed to form a disordered potential. We begin by identifying the relationship between the relative position of two point-like barriers and the critical velocity of such an arrangement. We then show that there is a mapping between the critical velocity of a system with two obstacles, and a system with a large number of obstacles. By establishing an initial superflow through a point-like disordered potential, moving faster than the critical velocity, we study how the superflow is arrested through the nucleation of vortices and the breakdown of superfluidity, a problem with interesting connections to quantum turbulence and coarsening. We calculate the vortex decay rate as the width of the barriers is increased, and show that vortex pinning becomes a more important effect for these larger barriers. 
\end{abstract}

\maketitle

\section{Introduction}
\label{sec-introduction}
A prototypical study of turbulence in fluids is that of the wake behind a cylinder in a flow \cite{Williamson1996a}. In classical fluids, the degree of turbulence in the flow can be encoded by the dimensionless Reynolds number $\mathrm{Re}=v D/\eta$, where $v$ is the velocity of the uniform flow, $D$ is the size of an obstacle in the flow, and $\eta$ is the kinematic viscosity of the fluid. Dynamical similarity allows us to map flows with different $v$, $D$ and $\eta$ to the same flow pattern, so long as the combination $vD/\eta$ is the same. In a superfluid flow, although $\eta \to 0$, it has been  shown that quantum fluids exhibit dynamic similarities in the same way classical fluids do \cite{Billam_PRL}.

A superfluid is characterized by frictionless flow in the absence of viscous effects. For a sufficiently small velocity, the flow around an obstacle is steady laminar flow and no vortices are nucleated \cite{Sasaki2010}. Above a critical velocity, the flow around an obstacle creates a drag force which is responsible for the nucleation of quantized vortices \cite{Frisch1992,Jackson1998}. These vortices signal the breakdown of superfluidity in the system at zero temperature \cite{Frisch1992,Winiecki2000,Inouye2001,Barenghi2008,Neely2010,Moon2015}. Immediately above the critical velocity,  pairs of oppositely charged vortices are shed periodically from opposite sides of the obstacle \cite{Kwon2015b}. As the velocity of the flow around the obstacle increases, there is a transition from the regular shedding of vortex dipole pairs to an irregular shedding of larger clusters of same-sign vortices, indicating that the system has become turbulent \cite{Sasaki2010,Kwon2016}. The transition to turbulence in superfluid flow past a potential obstacle has been the focus of recent theoretical \cite{Sasaki2010,reeves2013inverse,stagg2014quantum,Billam_PRL,stagg2016critical,Stagg2016_turbulence}  and experimental \cite{Kwon2014,Kwon2015a,Kwon2015b,Kwon2016} work. These works have investigated the effect of obstacle shape \cite{Winiecki1999,stagg2014quantum,Proment2019}  and finite temperature effects \cite{stagg2016critical,Stagg2016_turbulence} on the critical velocity for vortex nucleation past a single obstacle.  

As no real system is truly free of imperfections, disorder is an important consideration in interacting Bose systems, with the interplay between disorder and particle-particle interactions providing a rich test bed for many-body quantum physics.  Studies into disorder in BECs have employed impurities \cite{Giuriato2019}, rough boundaries \cite{stagg2017superfluid,Keepfer2020} and optical speckle patterns \cite{Clement2006,Pilati2009,Pilati2010,Pezze2011,Bourdel2012,Jendrezejewski2012,Krinner2013,Carleo2013,Cherroret2015,Meldgin2016,Scoquart2020}; the latter playing a role in the prediction of a lowered superfluid transition temperature  in 2D \cite{Bourdel2012} and 3D \cite{Pilati2009,Pilati2010,Bourdel2012}, the realisation of Anderson localization \cite{Jendrezejewski2012}, and the transition to an exotic Bose glass \cite{Krinner2013,Meldgin2016}.  While disorder is an important consideration,  few studies have considered the case of a superfluid flow in the presence of a point-like disordered potential. Such a disorder potential is now experimentally realisable, as new optical techniques employing technologies such as digital micromirror devices (DMDs) allow experiments to have an unprecedented level of control in creating arbitrary shaped potentials \cite{Henderson2009,Gaunt2013,Gauthier2016}.

Forcing a quasi-two-dimensional (2D) superfluid through a disordered potential faster than the critical velocity is a process which injects vortices into the system. These then decay by a process of vortex--antivortex annihilation, which is similar to the coarsening process which takes place after a thermal quench. Such coarsening is a current topic in 2D Bose gases, with investigations into the phase-ordering kinetics of this process being performed in conservative \cite{Damle1996,Schole2012,Billam2014,Simula2014,Karl2017,Groszek2018,Groszek2021} and dissipative \cite{Billam2014,Simula2014,Karl2017,Groszek2018,Comaron2019,Groszek2021} situations, as well as in systems of binary BECs \cite{Hofmann2014}, spinor BECs \cite{Williamson2016,Williamson2017}, and exciton-polariton condensates \cite{Kulczykowski2017,Comaron2018,Gladilin2019,Mei2021}. Previous works on single-component Bose gases have conducted quenches by starting from non-equilibrium initial conditions which tend to rapidly seed an approximately isotropic distribution of vortex dipoles \cite{BerloffSvistunov2002,Stagg2016_turbulence,stagg2016critical,Baggaley2018,Groszek2021}, while other studies have imprinted a random distribution of vortices with unit charge \cite{Groszek2018,Groszek2020,Groszek2020} or multiple charges \cite{Karl2017}. Here, we also observe a system which transitions from a non-equilibrium state containing many vortices towards an eventual equilibrium state, and use an energy- and number-conserving description similar to the conservative studies mentioned above. Our system, however, has several key differences. Firstly, the vortex injection in our system is different; unlike the initial conditions discussed above, the vortices which are created by a series of barriers have an anisotropic initial position which depends on the details of the barriers and the flow velocity of the superfluid. Secondly, the vortex injection is not instantaneous; rather vortices are shed over time from the barrier as the barrier moves through the superfluid above the critical velocity. Despite these differences, the system we describe provides a relatively simple way to generate non-equilibrium conditions which can be used to study related coarsening behaviour in a BEC.

In this paper we investigate the dynamics of dense 2D superfluid flow through a point-like disorder potential: a scenario which combines disorder, turbulence and coarsening in a 2D Bose gas. We impose the point-like disorder through an external trapping potential which is taken to be zero everywhere, apart from at a series of points where a localised repulsive barrier is placed. These repulsive barriers, which are Gaussian in shape, may be thought of as a set of blue-detuned laser beams whose intensity can be controlled at any point in space \cite{Henderson2009,Gaunt2013,Gauthier2016}. Unlike the disorder which is imposed by an optical speckle pattern, a key feature of this work is that the barriers which comprise the disorder potential are sufficiently separated (i.e. several healing lengths apart) so that the fluid is homogeneous away from the centre of the barrier. This ensures that it is possible to have a global superfluid phase, since localization of the condensate does not play a role \cite{Jendrezejewski2012}, and we can treat quantities such as the speed of sound and the healing length as effectively constant across the system.

The rest of the paper is structured as follows. In Sec.~\ref{sec-eom}, we describe the system and its equations of motion. 
In Sec.~\ref{sec-crit_vel}, we calculate the critical velocity for vortex nucleation for different point-like potentials. We begin by placing two identical point-like barriers in a superfluid flow, and study the interplay between relative separation and the incident angle of the barriers  on the critical velocity. We then look at a system with many point like barriers, and investigate the link between the density of these point-like barriers and the critical velocity of the system. 
In Sec.~\ref{sec-ringdown}, we study the long term behaviour of an initially non-equilibrium superfluid flowing through a disordered potential at varying initial velocities. We measure the condensate fraction, the superfluid fraction, and the superfluid velocity during this process. This illustrates how at short times the superflow breaks down, accompanied by vortex generation and depletion of the condensate fraction. At intermediate times, the momentum of the Bose gas continues to be arrested by interaction with the barriers, as vortex--antivortex annihilation begins. Over longer times vortices continue to annihilate and thermalization takes place; the gas recondenses and superfluidity is restored. 
In Sec.~\ref{sec-varying_barrier_width}, we investigate the effect of varying the effective barrier width on the vortex decay rate. For small point-like barriers (radius on the order of the healing length), the vortex decay rate follows the expectation for a thermal quench. We show that for sufficiently large barriers this changes, and at the same time vortex pinning becomes an important effect in the dynamics of the system. Sec.~\ref{sec-conclusions} contains our conclusions.

\section{System and Numerical Implementation}
\label{sec-eom}

We consider an obstacle which is moving at a steady velocity $\vv$ through a superfluid which is otherwise uniform in the $xy$ plane, and trapped strongly enough in the $z$ direction that all excitations are suppressed in this direction. Such a 2D system, when comprised of a weakly interacting atomic Bose gas at finite temperature, can be described by a wavefunction $\Psi$ which obeys the projected Gross-Pitaevskii equation, PGPE, 
\begin{equation}
    i \hbar \frac{\partial \Psi}{\partial t} =  \mathcal{P} \left\{ \left[ - \frac{\hbar^2}{2m} \nabla^2 + V_{\textrm{obj}}(\rr) + g_\mathrm{2D} |\Psi|^2 - \mu_\mathrm{2D} \right] \Psi \right\}.
    \label{eqn_dGPE}
\end{equation}
Here, $\mu_\mathrm{2D}$ is the chemical potential and the strength of the atomic interactions is parameterized by $g_\mathrm{2D} = \sqrt{8 \pi} \hbar^2 a_s / m l_z$, where $m$ is the atomic mass, $a_s$ is the $s$-wave scattering length, and $l_z = \sqrt{\hbar/m\omega_z}$ is the harmonic oscillator length in the $z$ direction. We impose a uniform flow with velocity $\vv$ in the $\boldsymbol{\hat{x}}$ direction by multiplying the initial wavefunction by a phase gradient (see, for example, Ref.~\cite{Winiecki1999}). The crucial feature of the PGPE, beyond the ordinary non-projected Gross-Pitaevskii equation, is the projection operator $\mathcal{P}$ which implements an energy cutoff in the basis of non-interacting single particle modes. When working at finite temperature, this allows one to set the cutoff so that modes below the cutoff are highly occupied. In this regime quantum fluctuations are relatively small and the classical field description is accurate \cite{Blakie_review}.

Alternatively, we can consider the system in which the obstacles are dragged through the fluid at some velocity $\vv$. In this system, the coordinate of the obstacle reference frame is $\rr = \rr_L + \vv t$, and the lab-frame wavefunction $\Psi(\rr,t) = \Psi_L \left( \rr_L, t \right)$. The PGPE governing the lab-frame wavefunction is given by
\begin{equation}
    i \hbar \frac{\partial \Psi}{\partial t} = \mathcal{P} \left\{ \left[ - \frac{\hbar^2}{2m} \nabla^2 + V_{\textrm{obj}}\left(\rr\right) + g_\mathrm{2D}|\Psi|^2 - \vv\cdot\pp - \mu_\mathrm{2D} \right] \Psi \right\},
    \label{moving_gpe}
\end{equation}
where the Gallilean shift to the obstacle frame (from the lab-frame) is given by the $\vv\cdot\pp$ term, with  $\pp = -i\hbar \nabla$ the usual quantum momentum operator \cite{Leadbeater2003,FFT_footnote}. 

To simulate an obstacle which is a collection of point-like barriers, we use the sum of $N_B$ repulsive Gaussian potentials,
\begin{equation}
    V_{\mathrm{obj}} (\rr) = V_0 \sum_{k=0}^{N_B} \exp \left[ - \frac{\left(x-x_k\right)^2}{a^2} - \frac{\left( y-y_k \right)^2}{a^2} \right],
\end{equation}
which have their centers at $\left(x_k, y_k\right)$. These barriers each have an effective cylinder width which can be estimated from the zero density region of the Thomas-Fermi approximation, $2a\sqrt{\ln \left(V_0 / \mu_\mathrm{2D} \right)}$. In contrast to previous works  which use hard-walled barriers \cite{Sasaki2010, stagg2014quantum}, we use soft-walled barriers (with $V_0 = e \mu_\mathrm{2D}$) and  where the critical velocity is lower \cite{Winiecki1999}. Unless otherwise stated, we take barriers to have a narrow waist, $a=\xi$, thus providing a point like potential with an effective cylinder width $2\xi$.

In what follows, we take $\vv = -v_{\mathrm{obst}} \boldsymbol{\hat{x}}$. The single-particle modes for this system are plane waves satisfying $|\kk|<k_\mathrm{cut}$, for some wave-number cutoff $k_\mathrm{cut}$. To implement the cut-off, we ignore $V_{\mathrm{obj}}$ as it will not affect the potential on the scale of $1/k_\mathrm{cut}$, and hence does not affect our choice of basis functions. The PGPE is evolved numerically, with doubly periodic boundary conditions, using an adaptive Runge-Kutta method (implemented using XMDS2 \cite{XMDS}) on a $L_x \times L_y$ grid with $N_x\times N_y$ grid points. We take the energy cutoff to be $k_\mathrm{cut} = \pi N_x / \left( 2 L_x \right) - \pi/L_x$. In the rest of the paper, we typically express quantities with reference to energy $\mu_\mathrm{2D}$, healing length, $\xi = \hbar / \sqrt{m \mu_\mathrm{2D}}$, density, $\rho=\mu_\mathrm{2D}/g_\mathrm{2D}$, and the speed of sound, $c=\sqrt{\mu_\mathrm{2D}/m}$. Consequently, times are expressed in units of $\tau = \hbar/\mu_\mathrm{2D}$. The number of grid points in our simulations is chosen such that there are two computational grid points per healing length.

\section{Critical Velocity of Point-Like Disordered Potentials}
\label{sec-crit_vel}

\subsection{Method}
In order to find the critical velocity, we first find the ground state of the condensate in the presence of the point-like potentials. To do this, we evolve the damped PGPE, found by multiplying the right hand side of Eqn.~\eqref{eqn_dGPE} by $(1-i\gamma)$, where $\gamma$ is a phenomenological damping parameter \cite{Choi1998_damping}, with stationary barriers $v_{\mathrm{obst}}=0$, and for $\gamma=1$, up to $t=5000\tau$. This converges to a wavefunction which is approximately the ground state of the system, and which will be the initial condition for all of the following simulations. We then set $\gamma=0$ and evolve Eqn.~\eqref{moving_gpe} , whilst smoothly ramping up the velocity \cite{stagg2014quantum} according to 
\begin{equation}
    \vobst (t) = v_{f} \tanh \left( \frac{t}{200 \tau}\right).
    \label{tanh_profile}
\end{equation} 
Smoothly increasing the velocity in this way prevents the generation of sound which would be caused by instantaneously setting $\vobst=v_f$. This simulation is run for $1000\tau$. The value of $v_f$ is increased discretely in small increments until vortices are observed to be shed from the potential. For reference, the critical velocity of a single point-like barrier is $\vcrit/c = 0.5625 \pm 0.0025$.

\subsection{A Pair of Point-like Barriers}
\label{sec-pairs}
\begin{figure}
    \centering
    \includegraphics{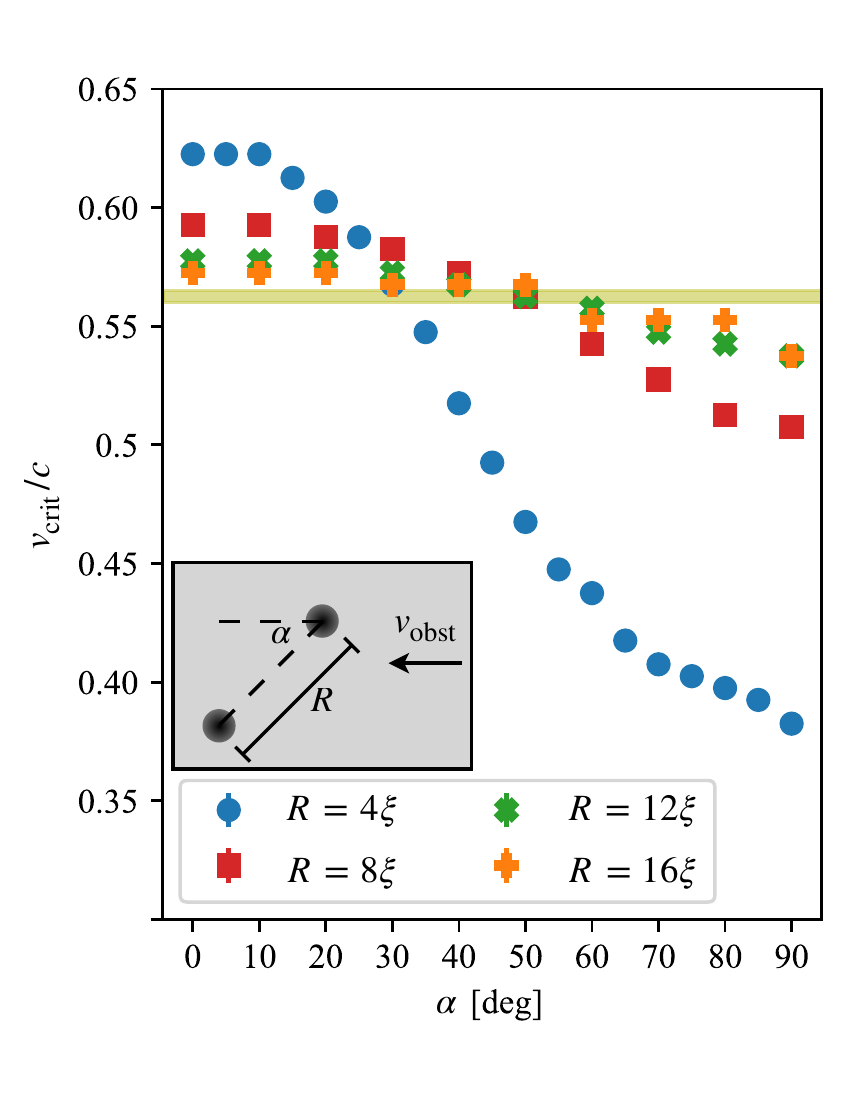}
    \caption{The critical velocity of two point-like barriers, with separation distance $R$, and angle $\alpha$ incident to the direction of the flow. Blue circles represent barriers with separation $R=4\xi$, red squares represent barriers with separation $R=8\xi$, green crosses represent barriers with separation $R=12\xi$, and orange pluses represent barriers with separation $R=16\xi$. The olive region is the critical velocity of a single point-like barrier, plotted as a guide to the eye (the width indicates numerical uncertainty). The error from the systematic uncertainty of increasing $v_x$ in discrete steps is smaller than the symbols used. The inset shows a schematic of the experimental set up of the two point-like barriers.} 
    \label{fig:pair_critical_velocity}
\end{figure}

We begin by finding the critical velocity of two point-like barriers, as we vary the relative distance and angle between these barriers. Without loss of generality, we place one barrier at the origin, and one barrier at $ \left( -R \cos \alpha, - R \sin \alpha \right)$. The results of this are plotted in Fig.~\ref{fig:pair_critical_velocity}.

When $\alpha$, the angle between the barriers in the direction of the flow, is small, the system has an increased critical velocity as the barriers are behind each other in the direction of the flow, becoming streamlined. As $\alpha$ increases, the critical velocity decreases since the barriers become a more like an effective elliptical obstacle, causing a denser wake \cite{stagg2014quantum}. An important observation that we make is that in the case where $R=4\xi$ the two barriers act as one larger (essentially elliptical) barrier, and for $v\gtrapprox\vcrit$ will shed only one dipole pair of vortices. In the cases where $R\geq 8\xi$, the barriers act independently and both of the point-like potentials will emit a dipole pair, for flow speeds just above $\vcrit$. The flattening of the curves indicates that, as we would expect, $\vcrit$ tends towards the single barrier result as $R\to \infty$.

\subsection{Multiple Barriers}
\label{sec-disorder}

\begin{figure}
    \centering
    \includegraphics{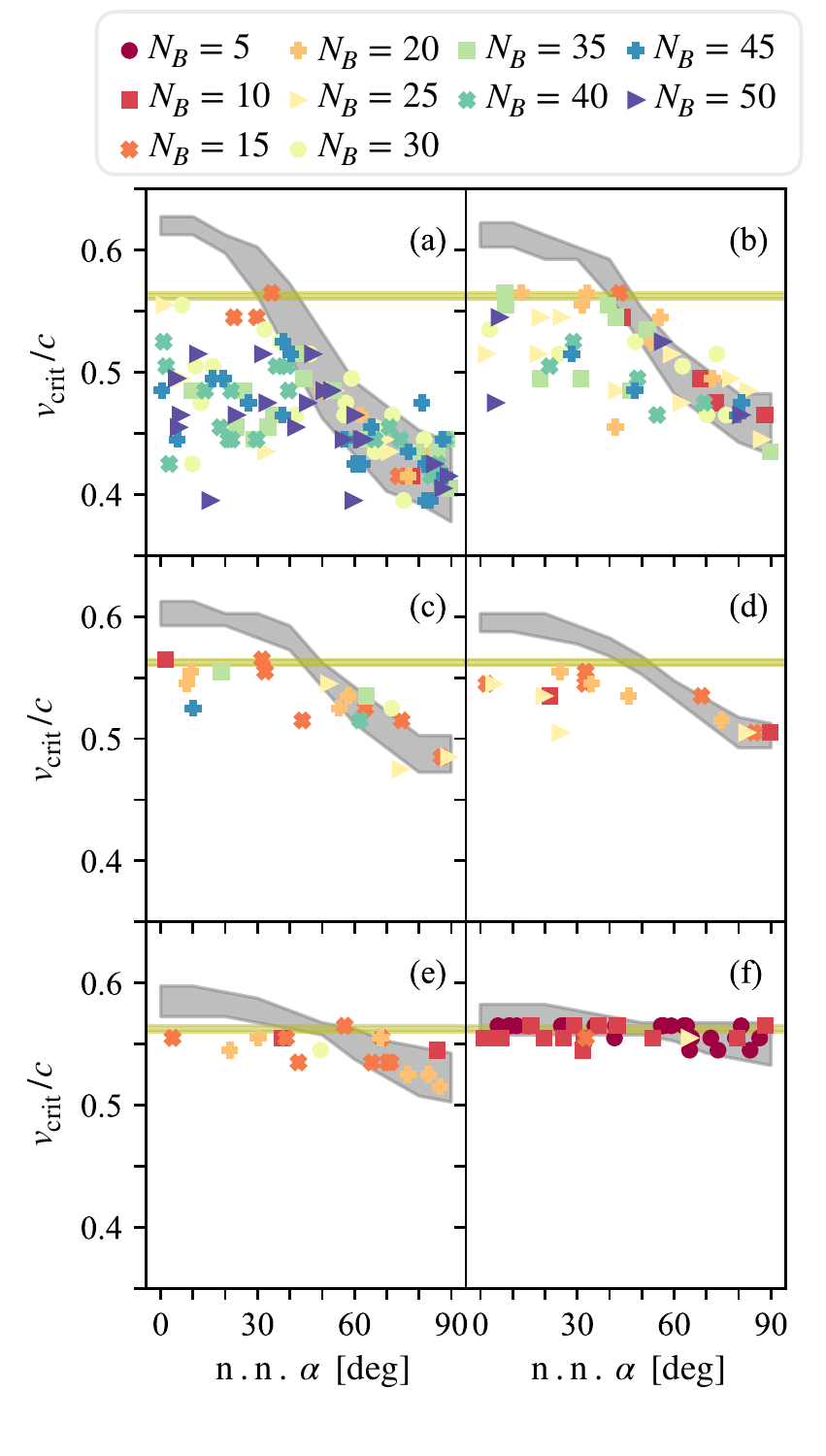}
    \caption{The critical velocity of a disordered potential with $N_B$ point-like barriers. Organised by nearest neighbour distance ($\mathrm{n.n.} \ R$) between the barriers, panel (a) has $4\xi \leq\mathrm{n.n.} \ R< 5\xi$, (b) has $5\xi \leq\mathrm{n.n.} \ R< 6\xi$, (c) has $6\xi \leq\mathrm{n.n.} \ R< 7\xi$, (d) has $7\xi \leq\mathrm{n.n.} \ R< 8\xi$, (e) has $8\xi \leq\mathrm{n.n.} \ R< 12\xi$, (f) has $12\xi \leq\mathrm{n.n.} \ R$.  Different markers represent varying barrier density.  The gray shaded area indicates the region containing the critical velocity of an isolated pair of point-like barriers when their separation distance lies within the range of nearest neighbour distances for the panel. The olive shaded area is the critical velocity of one point-like barrier (the width indicates numerical uncertainty).}
    \label{fig:disorder_critical_velocity.pdf}
\end{figure}

Having found the critical velocity for a pair of point-like barriers, we now find the critical velocity for $N_B$ barriers which are placed at random in the cell, subject to a minimum separation of $4\xi$.  A vortex detection algorithm similar to Ref.~\cite{Foster2010} is used to automate the search. 

In Fig.~\ref{fig:disorder_critical_velocity.pdf} we plot the critical velocity of a disordered system, as a function of the angle between the nearest neighbour pair of point-like potentials in the particular disorder realization, $\textrm{n.n.} \ \alpha$. We choose this measure because we anticipate that the critical velocity of a particular potential will be most sensitive to the configuration of the pair of barriers with the smallest separation, as shown by the range of values in the blue curve of Fig.~\ref{fig:pair_critical_velocity}.  The panels of Fig.~\ref{fig:disorder_critical_velocity.pdf} correspond to the binning of all realizations according to the nearest neighbour distance between the closest two point-like barriers in each realization, while the type of marker represents the total number of barriers in the system, $N_B$. The gray shaded area indicates the region which contains the critical velocity of a system of 2 point-like barriers whose separation distance corresponds to the separation distance of the panel. For larger $\textrm{n.n.}\alpha$, the nearest neighbour interactions of the closest pair of point-like barriers dominate the critical velocity, as can be seen by the points lying within the gray shaded region. Where the closest nearest neighbour barriers form a streamlined barrier, given by smaller $\textrm{n.n.}\alpha$, the critical velocity is smaller than the two barrier case; this is due to two factors. Firstly as $N_B$ increases, so does the probability that other (non-closest) pairs of nearest-neighbour barriers are separated by a similar distance but have a large angle against the flow, creating an efficient vortex emitter. Secondly, given that there are multiple barriers in the system, the critical velocity is limited by the single barrier case -- any barrier which is sufficiently separated ($\gtrapprox 20\xi$) from the other barriers will act independently, and cause vortices to be present in the system as soon as the flow velocity is greater than the critical velocity for a single point-like barrier. Indeed, we observe that the critical velocity of a point-like disordered potential is bounded above by the lowest of: (a) the critical velocity of a single barrier; (b) the highest critical velocity of the two barrier test case for equivalent nearest-neighbour separation of the closest two barriers.

\section{Arrest of a Superflow: Velocity Dependence}
\label{sec-ringdown}
\begin{figure*}
    \centering
    \includegraphics[width=7in]{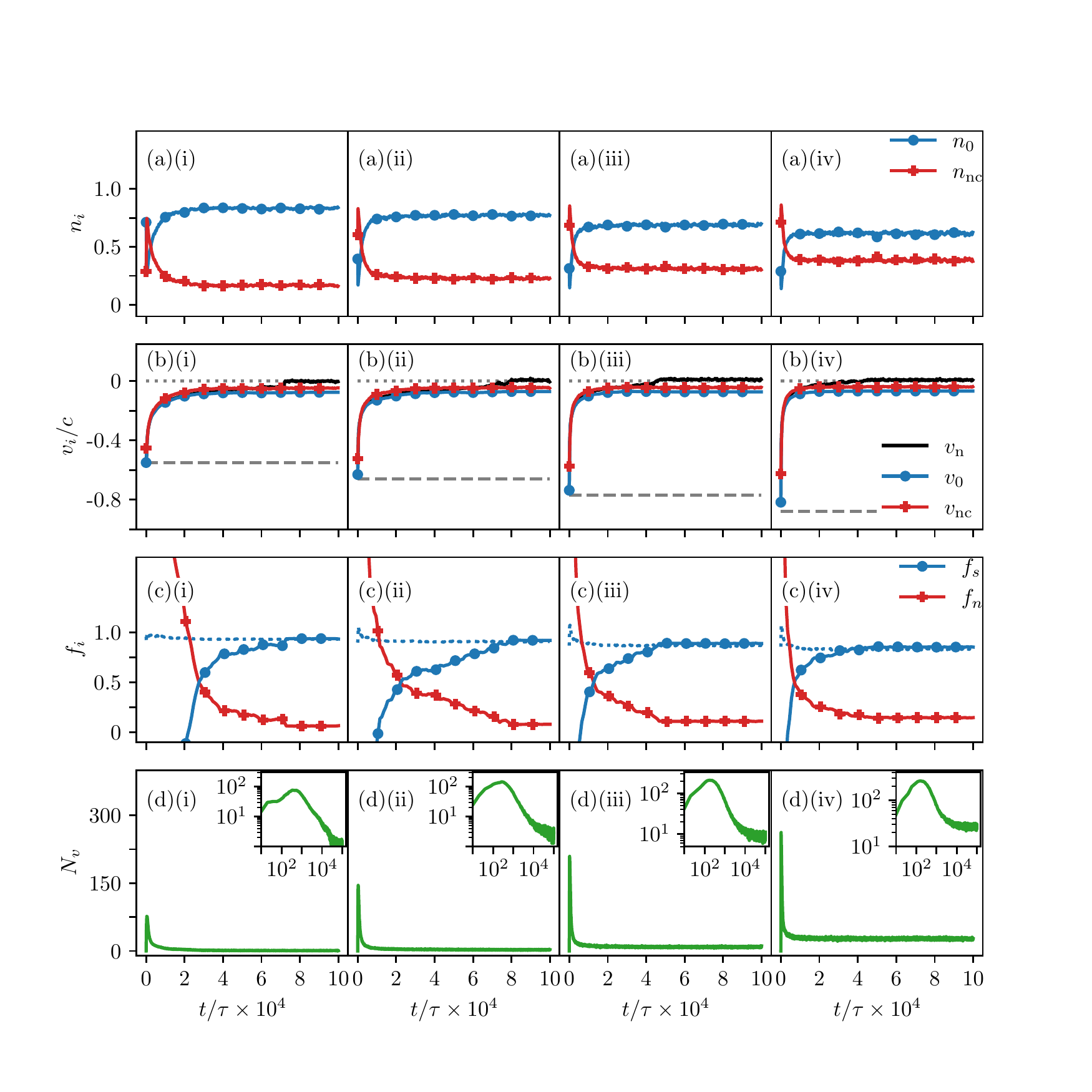}
    \caption{Evolution of statistics at different obstacle speeds for column (i) $\vobst=\vcrit$, column (ii) $\vobst=1.2\vcrit$, column (iii) $\vobst=1.4\vcrit$, and column (iv) $\vobst=1.6\vcrit$. Row (a) displays the condensate (blue circles) and non-condensate (red pluses) fractions. Row (b) is the velocity of the condensate mode (blue circles), the velocity of the the non-condensate mode (red pluses), and the approximate velocity of the normal fluid (black curve) given by Eqn.~\eqref{eqn:approx_normal_fluid_velocity}; the grey dashed line indicates $\vobst$, while the grey dotted line indicates zero velocity. Row (c) shows the superfluid fraction computed using the current--current correlations (blue circles), the approximated superfluid fraction described in the text (blue dotted line), and the normal fluid fraction (red pluses). Row (d) plots the vortex number; insets show the vortex number on a log-log scale.  The markers are added to help distinguish between curves, rather than indicating individual data points.  In the Supplemental Material \cite{SuppMat} we provide example movies of these simulations.}
    \label{fig:ringdown_statistics}
\end{figure*}

\subsection{Overview}

Driving a superfluid through a disordered potential faster than the critical velocity injects vortices into the system. The resulting non-equilibrium dynamics are a key object of study in two dimensional quantum turbulence, and have been employed as the initial conditions of studies into quenches both in the highly turbulent clustered case \cite{Karl2017}, and the dipole dominated case \cite{BerloffSvistunov2002,Stagg2016_turbulence,stagg2016critical,Baggaley2018,Groszek2021}. In this section, we consider a superfluid which is initially flowing through a disordered point-like potential, with an imposed velocity which is greater than the critical velocity of the potential. We observe that the reaction of the fluid is to be arrested by the barriers, with effectively viscous effects entering the system, before the system equilibrates.   The manner in which this disordered system reaches an equilibrium state has connections with quantum turbulence and coarsening in 2D Bose gases.

In this section, we consider one disordered potential with $N_B=25$ barriers in a system with dimensions $L_x = 256\xi$ by $L_y=64\xi$.  As the system consists of a superfluid initially moving through the point-like barriers above the critical velocity, the formation of elementary excitations causes the system to fall out of equilibrium \cite{Raman1999,Jackson2000}. To investigate the turbulence in such a system, we measure the condensate and non-condensate fractions, the velocity of the condensate and non-condensate fractions, the superfluid and normal fluid fractions, and the number of vortices which are nucleated by the obstacle. A similar setup has also been considered in  a 1D Bose gas subject to a series of randomly positioned delta scatterers \cite{Paul2007}, and more recently in a 2D Bose gas flowing through a blue-detuned speckle potential \cite{Cherroret2015,Scoquart2020}, however our random potential consists of point-like barriers between which there is a dense superfluid flow. 

In order to perform ensemble averaging we add a small amount of complex white noise to the groundstate of the wavefunction (with amplitude approximately equal to 1\% of the background density). This small amount of initial noise ensures that the system dynamics, and in particular vortex motion, differs in each realization, such that the statistics are not dominated by particular vortex trajectories. Averaging over this ensemble allows us to reliably calculate condensate fractions, condensate velocities, and superfluid fractions, as will be described in the following subsections.

In order to perform an analysis of this system in the long time limit,  we evolve the PGPE prescribed in Eqn.~\eqref{eqn_dGPE}. This is expressed in the frame where the barriers are at rest and the wavefunction is given an instantaneous initial boost,
\begin{equation}
    \Psi\left(\rr,0\right) = \left[ \left(1-\varpi\right) \Psi^{(g)}\left(\rr\right) + \varpi \varrho(\rr) e^{i\varphi(\rr)} \right] \exp\left(- \frac{2\pi i v_{\mathrm{int}} x}{c L_x} \right),
\end{equation}
where $\Psi^{(g)}(\rr)$ is the wavefunction in the ground-state of the system, $\varpi$ is the amount of noise to be added, $v_{\mathrm{int}}=\lceil \vobst / \Delta v\rceil \Delta v$, where $\lceil \cdot \rceil$ is the ceiling function, and $\Delta v = 2\pi c \xi / L_x$ is the smallest velocity representable on the grid in the $x$ direction. The random variables are $\varrho(\rr)\sim U[0,1]$ and $\varphi(\rr)\sim U[0,2\pi)$, and we renormalize such that the initial condition has the same normalisation as the ground state.
 We choose to evolve Eqn.~\eqref{eqn_dGPE} as it keeps the late-time, close-to-equilibrium momentum distribution of the system close to symmetric about $\kk=0$. Since the $k$-space cut-off imposed by the projector is symmetric about $\kk=0$, this choice ensures that the system is evolving towards a well-defined PGPE equilibrium, and hence that the calculation of the momentum-momentum correlations required to compute the superfluid fraction (Sec.~\ref{subsec_superfluid}) may be performed without the need to perform a gauge transformation.

\subsection{The Condensate and Non-Condensate Fractions}
\label{sec:cond_frac}

Where obstacles are dragged through a system at a speed sufficiently above $\vcrit$, a large number of vortex-antivortex pairs are nucleated, forming a complicated phase field \cite{Kwon2016}. Annihilation events between the vortex-antivortex pairs lead to the generation of sound in the system, which causes a depletion to the condensate fraction; this marks the onset of a dissipative regime.

Using the criterion of Penrose and Onsager \cite{Penrose1956}, within the c-field formalism \cite{Blakie_review}, we calculate the condensate and non-condensate fractions from the one body density matrix
\begin{equation}
    G^{1B}\left(\rr,\rr' \right) = \langle \Psi^* \left( \rr \right) \Psi \left( \rr' \right) \rangle_\mathcal{T}, 
    \label{one_body_density_matrix}
\end{equation}
where $\langle \ \cdot \ \rangle_\mathcal{T}$ indicates short time averaging. This fraction is calculated for each of the trajectories before averaging over all trajectories. The condensate number can be identified as the largest magnitude eigenvalue of the one-body density matrix, while the corresponding eigenvector, $\psi_0$, is the condensate mode. Under this formalism, we deconstruct the wavefunction into contributions from the condensate mode and a non-condensate mode, \begin{equation} 
\Psi = n_0 \psi_0 + n_{\mathrm{nc}}\psi_{\mathrm{nc}},
\end{equation}
where $n_0$ is the condensate fraction and $n_\mathrm{nc}$ is the non-condensate fraction, with $n_0 + n_{\mathrm{nc}} = 1$. The non-condensate mode, $\psi_\mathrm{nc}$, is the sum of the eigenvectors of $G^{1B}$ excluding $\psi_0$. Within the c-field formalism, the condensate and non-condensate modes are orthogonal.

The average condensate fractions for obstacles which are dragged through the fluid with velocity $v_{\mathrm{obst}} \geq \vcrit$ are plotted in Fig.~\ref{fig:ringdown_statistics}, row (a). In systems where $\vobst \geq \vcrit$, there is an initial depletion of the condensate fractions as the barrier sheds vortices which are subsequently annihilated, ultimately heating the system \cite{Neely2010}.  We observe that the size of the depletion of the condensate fraction (and therefore the spike in the non-condensate fraction) monotonically increases with the velocity of the obstacle which is consistent with the finding of Refs.~ \cite{Frisch1992,Winiecki1999}. Eventually, the energy that is injected into the system by annihilation events is distributed among phonons \cite{Frisch1992,Jackson2000}, and the system relaxes to a uniform flow. This is shown by the increase and then plateauing of the condensate fraction, indicating that the system has reached equilibrium and no further vortices are shed. We see that the long term behaviour of the condensate fraction depends on the speed of the obstacles; in the system where $\vobst=\vcrit$ the final condensate fraction is $n_0\approx0.84$,  where as for $\vobst=1.6\vcrit$ the final condensate fraction is $n_0 \approx 0.63$. This is to be expected, as the faster initial boost injects more energy into the system creating a hotter final state.

\subsection{The Velocity of the Condensate and Non-Condensate Modes}

As the shedding of vortices causes the depletion of the condensate fraction, we would expect the presence of thermal effects to lower the critical velocity \cite{stagg2016critical} which in turn would lead to the nucleation of more vortices, until the condensate is depleted. In fact, since the long term behaviour of the condensate fraction is to equilibriate, we deduce that the system stops shedding vortices. This indicates that the system dynamically reacts to the obstacle velocity. 

The velocity of the condensate mode $\psi_0$ (equivalently, the non-condensate mode $\psi_\mathrm{nc}$), is \cite{Pethick_and_Smith}
\begin{equation}
     \frac{\vv_k}{c} = \frac{1}{2i} \frac{\left( \psi_k^* \tilde{\nabla} \psi_k - \psi_k \tilde{\nabla} \psi_k^* \right)}{|\psi_k|^2},
    \label{sf_velocity}
\end{equation} 
where the index $k\in\{0,\mathrm{nc} \}$ and $\tilde{\nabla}=\xi\nabla$. We calculate the average velocities of the condensate, $v_0(t) = \left(L_x L_y\right)^{-1}\int d^2 \rr \ \vv_0\cdot\boldsymbol{\hat{x}}$, and non-condensate, $v_\textrm{nc}(t) = \left(L_x L_y\right)^{-1}\int d^2 \rr \ \vv_\textrm{nc}\cdot\boldsymbol{\hat{x}}$,   and plot them in the barrier reference frame in Fig.~\ref{fig:ringdown_statistics}, row (b).  

In the presence of the barriers, the fluid  nucleates vortices where $\vobst \geq \vcrit$. As these vortices nucleate, a phase winding is imparted on the wavefunction, which accelerates the fluid in an attempt to match the speed of the obstacle. In the long time limit, we observe that the velocity of the condensate and non-condensate modes is arrested by the barrier, suppressing further vortex nucleation. The drag force which is exerted on the obstacle potential by the fluid can also be measured \cite{Sasaki2010}, and we find that this vanishes as the system evolves. 

We note that we also expect to see a variation in the velocity in the $y$ direction, as different configurations of the barriers act as airfoils, \cite{Proment2019}, causing a lift effect. In our simulations, since $\vv = \vobst \boldsymbol{\hat{x}}$, this variation depends on the configuration of the barriers. In any realization it is small in comparison to the velocity change in the $x$ direction.

\subsection{The Superfluid and Normal Fluid Fractions}
\label{subsec_superfluid}

In order to understand the mechanism by which the velocity of the condensate and non-condensate modes are arrested by the barrier, we calculate the superfluid and normal fluid fractions. We calculate the superfluid fraction in two ways; further details of each approach are given in Appendix \ref{Appendix_superfluid}.

Firstly, we assume that the current, $\mathbf{J}$, of the wavefunction can be decomposed into contributions from a superfluid component (which flows without energy loss) and a normal fluid component (which is subject to viscous effects). We expect that the normal fluid will move with the barriers, and so in the frame of reference where the barriers are stationary, the velocity of the normal fluid will vanish in equilibrium. Since the superfluid velocity is locked to the condensate velocity \cite{Griffin,StatPhys2}, assuming zero normal fluid velocity leads to an estimate of the superfluid fraction $f_s$ using $\mathbf{J}=\rho f_s\mathbf{v}_0$, where $\mathbf{v}_0$ is the condensate velocity introduced in the previous section. 

Secondly, we compute the superfluid fraction by noting that the $(\alpha,\beta)$ element of the current-current correlations of the system in thermal equilibrium can be written as
\begin{equation}
   \left\langle \left\langle \left[\mathcal{F} \left( \mathbf{J} \right) \right]_\alpha \left[\mathcal{F} \left( \mathbf{J} \right) \right]_\beta^* \right\rangle_\mathcal{T} \right\rangle_\mathcal{R} \propto  f_s  \frac{k_\alpha k_\beta}{k^2} + f_n  \delta_{\alpha \beta}
   \label{eqn:current-current-propto}
\end{equation}
in the limit of vanishing momentum \cite{Pit_and_String}, where $\alpha,\beta \in \{x,y\}$. Here $f_s$ and $f_n$ are the superfluid and normal fluid fractions, and  $\mathcal{F}(\mathbf{J})$ is the Fourier transform of the current of the wavefunction. The angled brackets $\langle \langle  \ \cdot \ \rangle_\mathcal{T} \rangle_\mathcal{R}$ indicate that the correlations are found by short-time averaging and by averaging over the ensemble of initial conditions. It is then possible to extract the superfluid and normal fluid fractions by fitting the current-current correlations of the wavefunction to the right hand side of Eqn.~\eqref{eqn:current-current-propto} \cite{Foster2010}. Formally this method is only valid at equilibrium. Here we employ it with ensemble- and short-time-averaging to give a dynamic measure. We expect this measure to be quantitatively accurate at late times as equilibrium is approached, since we observe the current-current correlations are well fitted by the expected functional form of Eqn.~\eqref{eqn:current-current-propto} at later times. At earlier times, further from equilibrium, fits to the expected functional form of the correlations fail, and the measure only provides a qualitative indication of the lack of superfluidity.

The superfluid fraction of the system is plotted in Fig.~\ref{fig:ringdown_statistics}, row (c). Where the superfluid fraction (computed using the current--current correlations) is negative, or the normal fluid fraction is greater than 1, it is clear that the condition of vanishing momentum is not met. This condition is better fulfilled at later times, where the velocity of the fluid has been arrested by the barrier [see Fig.~\ref{fig:ringdown_statistics} row (b)]. The fluid must respond to the boost which is initially imposed, and so the velocity of the normal fluid at $t=0$ is not necessarily zero. This explains why, at very early times, the superfluid fraction computed by decomposing the momentum of the wavefunction is greater than 1.

 At earlier times, there is a jump in the normal fluid fraction which equates to the absence of superfluidity. It is this mechanism which causes the fluid to be arrested by the barriers: the appearance of many vortices is associated with a rise in the normal fluid component which is subject to viscous effects causing the fluid to be decelerated by the barrier. At later times, the superfluid fraction grows and equilibrates with the fluid velocity now approximately zero. By the end of the simulation, both measures of the superfluid fraction are close to each other. As the velocity of the barriers is increased, the final superfluid fraction decreases. In the system where $\vobst=\vcrit$ the superfluid fraction, averaged over the last 20\% of the simulation is $\bar{f_s}=0.94$ (with standard deviation $0.0009$) using current--current correlations, and $\bar{f_s}=0.93$ (with standard deviation $0.0015$) by decomposing the current of the wavefunction; in the system where $\vobst=1.6\vcrit$ these values are $\bar{f_s}=0.85$ (with standard deviation $0.0014$) and $\bar{f_s}=0.83$ (with standard deviation $0.0047$) respectively. This is an analogous result to the depletion of the final condensate fraction as $\vobst$ increases, as discussed in \secreft{sec:cond_frac}. 
 
 While a slow but non-zero final velocity of the superfluid (i.e., a slow remnant superflow) is not physically unexpected, it is interesting that we do not observe the non-condensate velocity $v_\mathrm{nc}$ reaching zero over the timescale of our simulations, as can be seen in Fig.~\ref{fig:ringdown_statistics} row (b). However, as noted above and as can be seen in Fig.~\ref{fig:ringdown_statistics} row (c), while the superfluid fractions found by current-current correlations and found by assuming that the momentum of the fluid is due entirely to the superflow are close, the two quantities are not equal. There is also a substantial difference between the superfluid and condensate fractions at late times. These observations suggest that part of the non-condensate fraction contributes to the superflow. This prompts a useful consistency check on our results: the expected total momentum of the fluid can be written in terms of either the superfluid and normal or the condensate and non-condensate components. Therefore, for the average velocities in the $x$-direction we should have 
  \begin{equation}
     f_s v_s + f_n v_n = n_0 v_0 + n_\mathrm{nc} v_\mathrm{nc},
     \label{eqn:momentum_eq}
 \end{equation}
 where $v_s$ and $v_n$ are the superfluid and normal fluid velocities. Assuming that the superfluid velocity is locked to the condensate velocity, $v_s = v_0$, but \textit{relaxing} the assumption that $v_n=0$, we can use our estimates of $n_0$, $n_\mathrm{nc}$, $v_0$ and $v_\mathrm{nc}$ from the Penrose--Onsager analysis and our estimates of $f_s$ and $f_n$ from the current-current correlation analysis to extract the normal fluid velocity from Eqn.~(\ref{eqn:momentum_eq}) as
  \begin{equation}
     v_n = \frac{\left(n_0 - f_s\right)v_0 + n_\mathrm{nc} v_\mathrm{nc}}{f_n}.
     \label{eqn:approx_normal_fluid_velocity}
 \end{equation}
 The value of normal fluid velocity $v_n$ obtained from Eqn.~(\ref{eqn:approx_normal_fluid_velocity}) is shown in Fig.~\ref{fig:ringdown_statistics} row (b). One can see at late times that the zero value $v_n=0$ is within the range of the time-variation in this quantity, and the average $v_n$ is much closer to zero than the average $v_\mathrm{nc}$. We expect the slight remaining offset of the average $v_n$ from zero results from the combination of the various statistical uncertainties in our simulations and the Penrose--Onsager and current-current correlation analysis that feed into Eqn.~(\ref{eqn:approx_normal_fluid_velocity}). Overall our simulations and analysis show a consistent picture that, over the timescale of our simulations, interaction with the barriers has resulted in a remnant superflow at well below the cricital velocity coexisting with a normal fluid component that has been slowed to very close to zero velocity with respect to the barriers.

\subsection{The Vortex Number}
Since the reaction of the fluid is to accelerate to catch up with the barriers, vortex--antivortex pairs are shed from the barrier only at the beginning of the simulation. This leads to a peak in the vortex number as seen in row~(d) of Fig.~\ref{fig:ringdown_statistics}. It is evident that the amplitude of the peak in $N_\mathrm{v}$ increases as $\vobst$ increases; this is because the vortex shedding frequency increases with the velocity of the obstacle \cite{Winiecki1999}. 

At the end of the simulation it is possible that a small number of vortices remain in the system. The average number of such vortices at late times increases as the late-time condensate and superfluid fractions decrease. Typically, for the barriers considered in this section, this small number of vortices are not pinned to barriers but are free to move, and hence consistent with thermal vortices in the fluid.  We will discuss the role of free vortices and vortices which become pinned to the barriers in more detail in the next section.

It should be emphasized that, while the results presented in this section are measurements of one disordered potential averaged over an ensemble of ten initial conditions, these results are applicable to other disordered potentials. We have checked that the results presented in Fig.~\ref{fig:ringdown_statistics} are qualitatively the same for other $N_B$, so long as $\vcrit$ is the same (within error bars). The effect of simulating a system which has a higher (lower) $\vcrit$ is simply to steepen (flatten) the curves seen in Fig.~\ref{fig:ringdown_statistics}, while the long-term behaviour is unchanged.

\section{Arrest of a Superflow: Scaling and Turbulence}
\label{sec-varying_barrier_width}

\begin{figure*}
 \centering
 \includegraphics[]{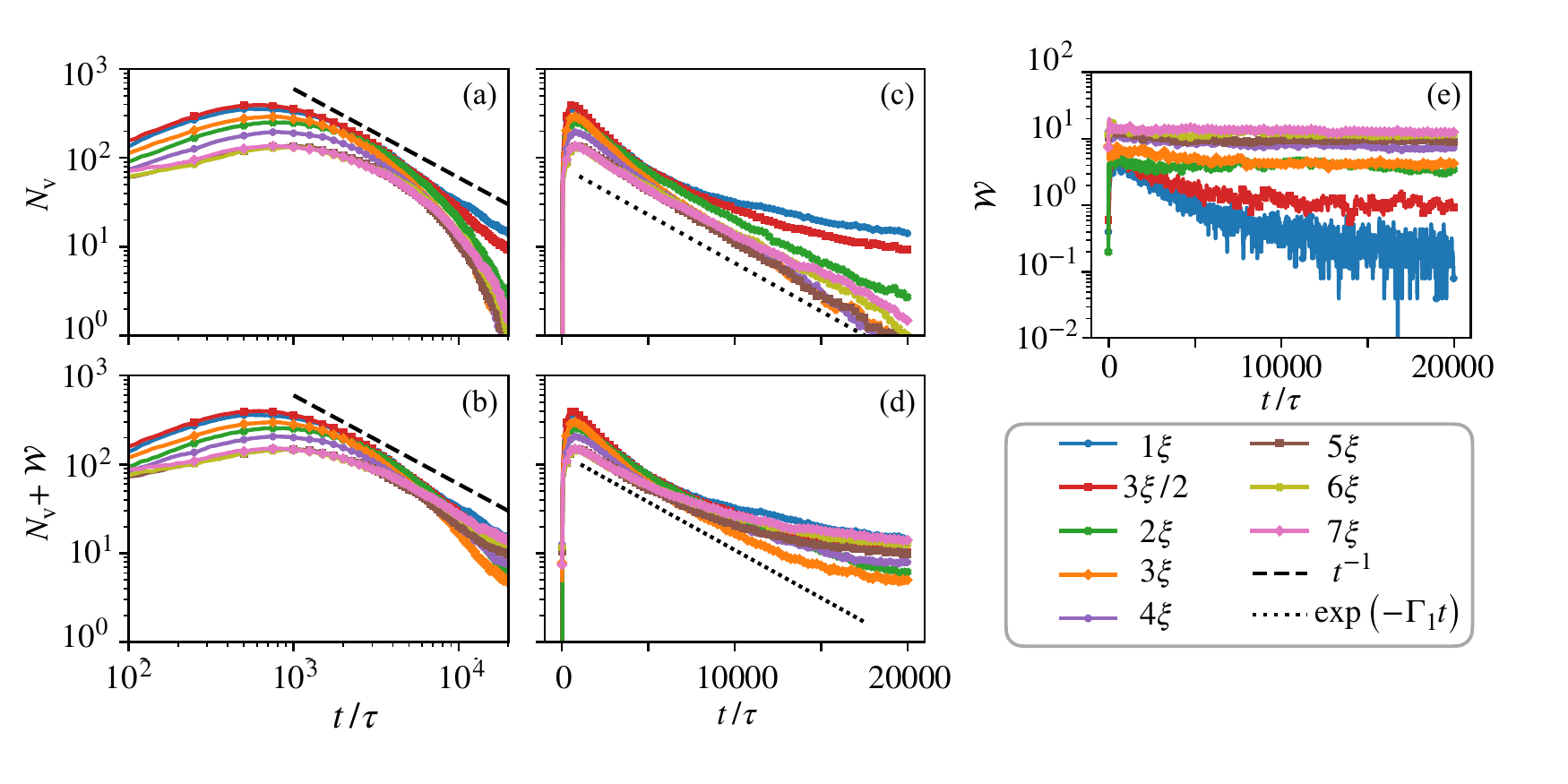}
 \caption{The decay of the number of vortices in a system as the barrier width varies. Panels (a) and (c), the decay of the number of mobile vortices, $N_v$. Panels (b) and (d), the decay of the total number of vortices, $\left(N_\mathrm{v}+\mathcal{W}\right)$. Panel (e), the decay of the number of pinned vortices, $\mathcal{W}$.   Panels (a) and (b) are plotted on a log-log scale, while panels (c)--(e) are plotted on a semi-log scale. The power law $N_\mathrm{v}\propto t^{-1}$, black dashed line, and the exponential decay $N_\mathrm{v}\propto \exp\left(-\Gamma_1 t\right)$, black dotted line, are added as guides to the eye. The markers are added to help distinguish between curves, rather than indicating individual data points.  }
 \label{fig:log_vortex_decay}
\end{figure*}

\subsection{Overview}
Until now we have only considered disordered potentials which consist of a number of point-like barriers, with an effective radius of $1\xi$, randomly placed in a periodic cell. We now extend our parameter space to consider disordered potentials consisting of barriers with a greater effective radius, and focus on analyzing vortex decay processes. In this section we consider a square domain with dimensions $L_x = L_y = 256\xi$.

The numerical simulations which are carried out in this section can be related to practical experiments. Periodic boundary conditions, such as those imposed in our simulations, can be realised in one direction in experiments using ring traps \cite{Ramanathan2011} It is possible to impose a persistent superflow current in such a geometry by stirring \cite{Eller2020} or optical methods \cite{Ramanathan2011}, creating a superflow in the periodic direction. Technology such as DMDs could be used to paint the stationary disordered potential in part or all of the ring trap \cite{Gauthier2016}. For a large, annular (i.e., tightly confined in the $z$-direction) ring trap, the main difference from our simulations here would be the lack of periodic boundary conditions perpendicular to the flow. We do not expect that difference to play a crucial role in the dynamics as long as the difference between inner and outer radii of the annulus is a large number of healing lengths. Interestingly, in addition to the studies performed here, in such a system one could switch off the disorder potential after the initial burst of vortex injections; this could be used as a controllable way to inject a vortex distribution and study the resulting coarsening dynamics \textit{without} the point-like disorder.

\subsection{Vortex Decay Rate}
The rate at which a distribution of vortex dipoles in a quasi-2D Bose gas decays has been the subject of much discussion over the last decade \cite{Schole2012,Kwon2014,Groszek2016,Cidrim2016,Baggaley2018}. The vortex decay rate is expected to be connected to the growth of the correlation length of a system, $L_c$. As the system relaxes after a quench, $L_c$ should become the only relevant length scale, and it is predicted that $L_c$ grows as $L_c(t)\sim t^{1/z}$, where $z$ is the dynamical critical exponent \cite{Bray_review_1994}. It is also predicted that, for randomly distributed vortices in a homogeneous system, the vortex number and the correlation length are linked as $N_\mathrm{v}\sim L_c^{-2}$. Based on experimental observations, the suggested phenomenological rate equation for $N_\mathrm{v}$ is \cite{Kwon2014}
\begin{equation}
    \frac{dN_\mathrm{v}}{dt} = - \Gamma_1 N_\mathrm{v} - \Gamma_2 N_\mathrm{v}^2.
    \label{eqn:Kwon_decay_rate}
\end{equation}
Single vortex annihilations are prohibited as vortices are topologically protected, meaning that $\Gamma_1 N_\mathrm{v}$ describes the drifting of vortices out of the condensate at \textit{boundaries} (a one-vortex mechanism), while $\Gamma_2 N_\mathrm{v}^2$ represents the rate of vortex-antivortex annihilations (a two-vortex mechanism, in this model). However, the decay rate given by Eqn.~\eqref{eqn:Kwon_decay_rate} does not match with the results of zero-temperature GPE simulations \cite{Schole2012,Cidrim2016,Groszek2016,Baggaley2018}. This has led to the proposal of a corrected idealized decay rate \cite{Cidrim2016}
\begin{equation}
    \frac{dN_\mathrm{v}}{dt} = -\Gamma_1 N_\mathrm{v}^{3/2} - \Gamma_4 N_\mathrm{v}^4,
\end{equation}
where it is argued that the drift and annihilation processes have a $N_\mathrm{v}^{3/2}$ and $N_\mathrm{v}^4$ dependence respectively. It has since been shown \cite{Cidrim2016,Groszek2016,Baggaley2018} that for a homogeneous system at zero temperature $N_\mathrm{v} \sim t^{-1/3}$ which is indicative of a four-vortex process, while the addition of dissipation (finite temperature effects) or trapping potentials removes the need for a fourth vortex \cite{Groszek2020} (the $N_\mathrm{v}^4$ scaling which describes a four-vortex annihilation process was also observed numerically in Ref.~\cite{Schole2012}).

Due to the large proportion of our simulations which occur after the peak in vortex number, it is possible to study the long-time behaviour of vortex decay in our disordered potential systems in a similar fashion. As discussed earlier, we use the plaquette technique \cite{Foster2010} to enable vortex detection. Unlike before, where we focused on barriers with effective radii $1\xi$, for barriers which have an effective radii $\gtrsim 2\xi$ there is a significant zero density region where the phase of the condensate is ill-defined. Naively applying the plaquette technique here leads to the detection of spurious vortices. However, it is also possible for a net number of quanta of circulation to genuinely be present at this low density region: we define this number of quanta as the \textit{winding number} of the barrier $\mathcal{W}_k$ (for the $k$th barrier). The winding number can also be interpreted as a number of \textit{pinned} vortices. Hence, when computing the vortex number we detect both the number of \textit{mobile} vortices $N_\mathrm{v}$, using the plaquette technique and excluding the density-depleted regions, and the total number of \textit{pinned} vortices
\begin{equation}
    \mathcal{W} = \sum_{k=1}^{N_\mathrm{v}} |\mathcal{W}_k|,
\end{equation}
which is computed using a loop integral technique described in the next section.

\begin{figure}
    \flushleft
    \includegraphics{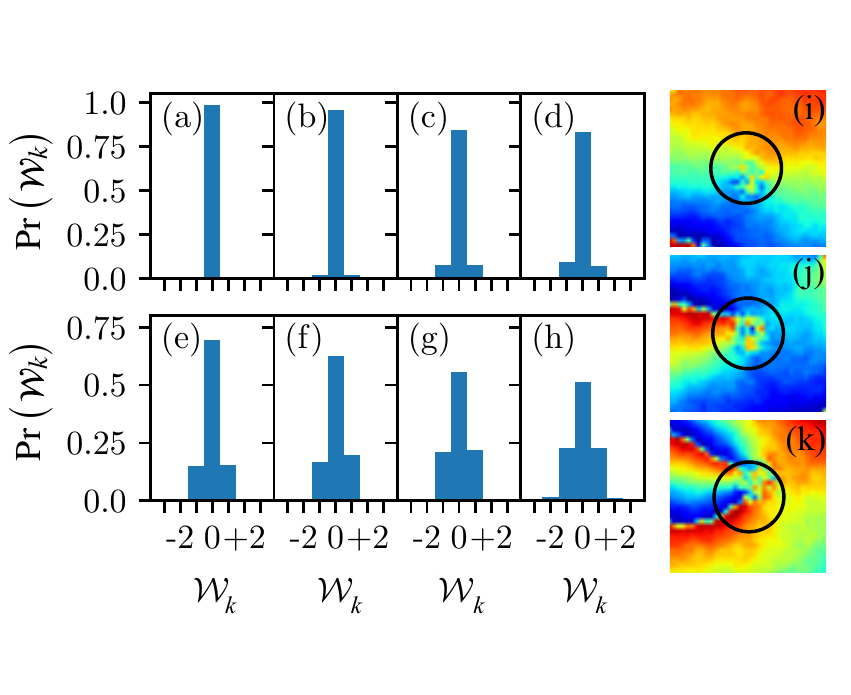}
    \caption{Normalized histogram of the winding number, $\mathcal{W}_k$, for barriers with effective radius (a) $1\xi$, (b) $3\xi/2$, (c) $2\xi$, (d) $3\xi$, (e) $4\xi$, (f) $5\xi$, (g) $6\xi$ and (h) $7\xi$. Right: examples the phase of the wavefunction, $\Arg\left(\Psi\right)$, around a barrier with (i) no pinning, $\mathcal{W}_k=0$, (j) one pinned vortex, $\mathcal{W}_k=+1$, and (k) two pinned vortices, $\mathcal{W}_k=+2$; the black circle is approximately the boundary of the zero-density region of the barrier. A slightly larger circular loop is used to compute the integral required to calculate $\mathcal{W}_k$.
    \label{fig:winding_histogram}}
\end{figure}

The evolution of the vortex numbers for a system with $N_B=25$ barriers of varying effective radii is shown in Fig.~\ref{fig:log_vortex_decay} (a)--(d). In Fig.~\ref{fig:log_vortex_decay}(a) and (c) we plot only the number of mobile vortices $N_\mathrm{v}$. In Fig.~\ref{fig:log_vortex_decay}(b) and (d) we plot the total number of vortices (mobile and pinned), $N_\mathrm{v} + \mathcal{W}$. For the narrowest barriers we consider, the vortex decay rate appears to follow a $t^{-1.1}$ power law for effective barrier radii of $\xi$, and a $t^{-1.2}$ power law for effective barrier radii of $3\xi/2$, as can be seen in panel (a). In a system where the vortex number only decays via vortex-antivortex annihilations, Eqn.~\eqref{eqn:Kwon_decay_rate} predicts that $N_\mathrm{v}\propto t^{-1}$.  The fact that the observed power laws are relatively close to $t^{-1}$ for the narrowest barriers is indicative of the fact that vortex decay is a two-vortex process in this system. For barriers which are larger than the typical size of a vortex core (i.e., have an effective width which is greater than a few healing lengths), the vortex number appears to decay exponentially, as can be seen in panel (c). This is consistent with a solution to Eqn.~\eqref{eqn:Kwon_decay_rate} where a one-vortex mechanism is dominant, i.e., $N_\mathrm{v}\propto \exp\left(-\Gamma_1 t\right)$. This suggests that for wider barriers, at late times in the simulation, vortices are colliding with a barrier more often than they are colliding and annihilating with a vortex of the opposite sign. We discuss the effects of vortices colliding with barriers in the following section.

\begin{figure}
    \centering
    \includegraphics{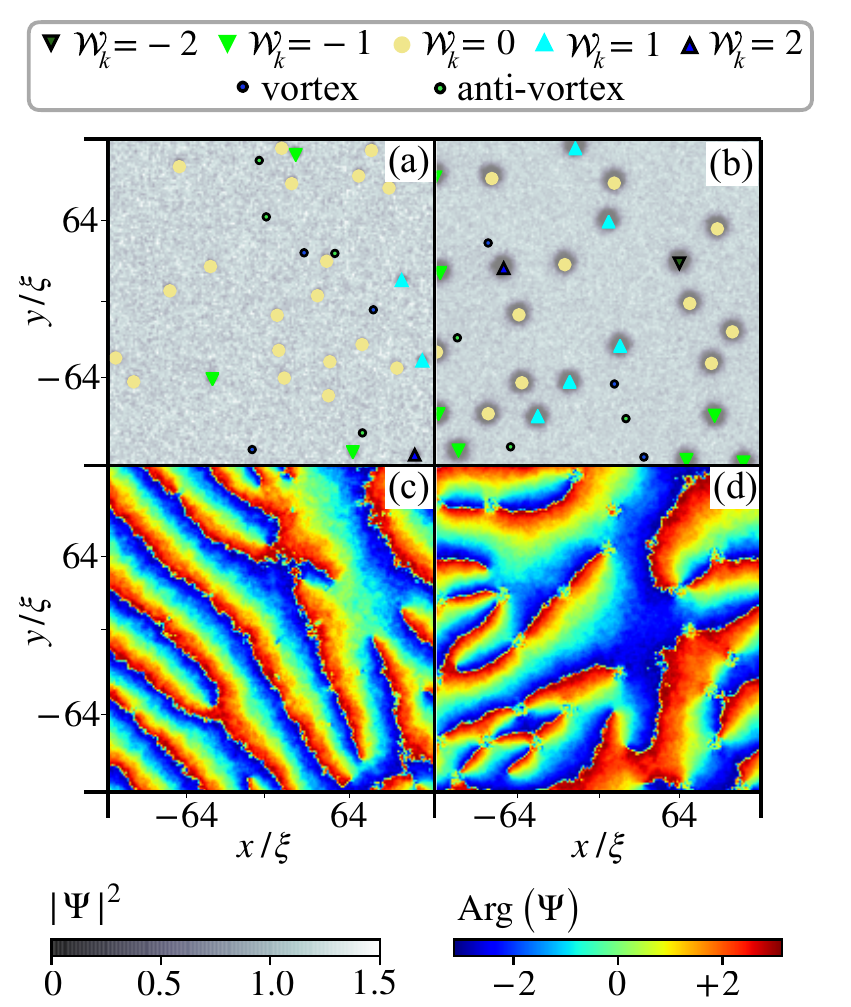}
    \caption{A still at time $t=14460\tau$ from the simulations of barriers with an effective radius of $3\xi$, (a) and (c), and simulations of barriers with an effective radius of $7\xi$, (b) and (d). In panels (a) and (b), the density of the wavefunction is shown, while different markers indicate the winding number $\mathcal{W}_k$ of a barrier, and the position of a vortex or anti-vortex. The phase of the wavefunction  is shown in panels (c) and (d). In the Supplemental Material \cite{SuppMat} we provide example movies of these simulations.}
    \label{fig:simulation_still}
\end{figure}

\subsection{Pinning to Barriers}
\label{subsec:pinning}
As well as measuring the rate at which the number of vortices decay, we have also measured the number of vortices which become pinned to the barriers. The pinning and un-pinning of superfluid vortices is an important physical process for understanding the mechanism of neutron star glitches \cite{Anderson1975,Jones1997,Jones1998,Donati2003,Link2009}, and is also of interest in systems with macroscopic container defects \cite{Schwarz1981pinning,DeBlasio1998,Jones2003}, as well as spin-down experiments with helium \cite{Tsakadze1975,Andronikashvili1979}, and laboratory BECs \cite{Bhat2006}. The microscopic process by which a vortex becomes pinned to a density depleted region has recently been studied by  Ref. \cite{Stockdale2021}. For systems where impurities exist, it is energetically favourable for a vortex to be contained within the zero-density region, as there is no cost in energy to create a vortex core \cite{Warszawski2012}.  

As described above, we define \textit{pinned} vortices in terms of the net quanta of circulation around a barrier, which is well defined as the branch cut representing a discontinuity in the phase extends into the non-zero density region of the condensate (i.e., it is not a spurious vortex caused by the phase not being well defined in the zero density region at the centre of the barrier). For each barrier in a given potential, we can measure the winding number $\mathcal{W}_k$ by integrating around a loop containing the barrier (see Appendix \ref{appendix_ring_plaquette} for details of the numerical method). Examples of the phase of a barrier with no pinned vortices, one pinned vortex, and two pinned vortices are shown in Fig.~\ref{fig:winding_histogram} panels (i), (j) and (k). Also shown is the approximate location of the radius of the circular ``exclusion zone'' which we choose when counting the number of mobile vortices. A slightly larger circular loop is used to measure the winding number. It should be noted that in any one trajectory the time-dependent values of the numbers of mobile and pinned vortices may display fluctuations in time that depend on the precise choice of radii for these circles, especially when two or more barriers are close together. While we were unable to find choices that eliminate these fluctuations in any one trajectory, we find the averaged results are relatively insensitive to the choice of radii.

At early times, the system is in a highly non-equilibrium state, and many vortices are periodically shed by the barriers. However, by $t\gtrapprox10^3\tau$, shedding from each of the barriers has almost completely stopped, and the winding number of each barrier is steady. This can be seen in Fig.~\ref{fig:log_vortex_decay} panel (e). For larger barriers, the number of mobile vortices in the system decays as $N_\mathrm{v}\propto \exp\left(-\Gamma_1 t\right)$, suggesting that the vortices are annihilating with the barriers. From our observations of the simulations, we suggest that there are 3 processes taking place here. Process I: a vortex collides with a barrier which has a number of vortices with the same sign pinned to it. Here the number of mobile vortices decays, $N_\mathrm{v}\to N_\mathrm{v}-1$, while the number of pinned vortices increases, $\mathcal{W}\to\mathcal{W}+1$. Process II: a vortex collides with a barrier which has a number of vortices with the opposite sign pinned to it. Here the number of mobile vortices decays, $N_\mathrm{v}\to N_\mathrm{v}-1$, but the number of pinned vortices also decays since the mobile vortex annihilates with one of the pinned vortices, $\mathcal{W}\to\mathcal{W}-1$. Process III: a dipole pair collides with a barrier which has a number of vortices pinned to it. Here, the number of mobile vortices decreases by two, $N_\mathrm{v}\to N_\mathrm{v}-2$, however the number of pinned vortices remains the same, since one of the dipole pair will annihilate with the vortices of opposite sign in the barrier, while the other vortex in the dipole pair will remain and will become pinned to the barrier, $\mathcal{W}\to\mathcal{W}$. As each barrier sheds an equal number of vortices and anti-vortices, Processes I and II take place with approximately the same frequency, conserving the pinning number $\mathcal{W}$. Process III, which also conserves the winding number, happens far less frequently. However, this process may perhaps explain the slight modifications to the exponential decay which we see in Fig.~\ref{fig:log_vortex_decay}. We assume that collisions between three or more vortices and a barrier are so rare as to be negligible.

The probability of observing a given winding number can be seen in the histograms in Fig.~\ref{fig:winding_histogram}, where the data is taken from $10^4\tau \leq t \leq 2\times10^4\tau$. As we can see, for narrow barriers vortex pinning is not an important feature. However, for barriers which are significantly larger than a vortex core, a significant number of the barriers do have a vortex or anti-vortex pinned to them $(\mathcal{W}_k=\pm1)$, and the largest barriers which we consider support the pinning of multiple vortices $\left(|\mathcal{W}_k|>1\right)$. Examples of this behaviour can be seen in Fig.~\ref{fig:simulation_still}.

It can be seen in Fig.~\ref{fig:log_vortex_decay} that the rate at which the number of mobile vortices decays becomes quicker as the effective radius is increased past $2\xi$, and is at its  fastest for barriers which have an effective radius of $ \approx 5\xi$. This may be attributed to the fact that for  barriers with an effective radius greater than $2\xi$ we have observed that is is more likely for a barrier to support the pinning of vortices; this provides a mechanism to lose mobile vortices via Process I, above. For barriers which have a larger effective radius than $5\xi$, we have observed that it is possible to have multiple vortices pinned to a barrier. Multiple pinning creates a stronger velocity field around the barrier than single pinning does; this could explain why the rate at which the number of mobile vortices decays slows slightly as the effective barrier radius increases above $5\xi$.

\section{Conclusion}
\label{sec-conclusions}

In this paper we have studied the effect of dragging a disordered point-like potential through a superfluid which is initially in the ground state. We have seen how the critical velocity of two point like barriers depends on the relative distance and angle between the barriers. We have then determined the critical velocity for a system which has up to 50 point like barriers at randomized locations, and shown that the critical velocity of such a system can be mapped on to the two-barrier case by considering the separation and angle with respect to the flow of the closest nearest-neighbour pair of barriers in the disorder potential. 

Using PGPE simulations, we investigated the evolution of a system in which an initial superflow, moving at or above the critical velocity, is disturbed by a stationary point-like disorder potential. This strongly non-equilibrium initial condition causes the nucleation of vortices and depletion of the condensate and superfluid fractions. We observe that the reaction of the fluid is to accelerate to a final velocity closer to the obstacle velocity. This suppresses the nucleation of further vortices, and the fluid re-condenses and some superfluidity is restored.  

We extended our parameter space to consider the effect of larger barriers in the system, and investigated the way in which this affects the decay of the number of vortices in the system. It is clear that the presence of randomly placed barriers that have an effective width which is larger than the characteristic size of a vortex core, modifies the form of the vortex number decay from the behaviour identified in previous theoretical works without a disordered potential. Within the limits of our numerical analysis, it appears as though the vortex decay rate no longer follows a $t^{-1}$ power-law scaling which is indicative of vortex-antivortex annihilations, but rather the vortices collide with the barriers which make up the potential, causing an exponential  decay. This one-vortex decay process is confirmed with our observations of the simulations.  Finally, we observe that for these larger barriers vortex pinning becomes a relevant phenomenon, with the largest barriers which we consider supporting the pinning of multiple vortices. 

With an appropriate trapping geometry, it may be possible to experimentally study a system equivalent to the one studied here in which the disordered potential arrests a superflow.

\acknowledgments
We thank George Stagg for helpful discussions whilst preparing this manuscript. This research was supported by  the UK EPSRC [Grant Nos. EP/N509528/1 and EP/R021074/1], the Australian Research Council Centre of Excellence in Future Low-Energy Electronics Technologies [Project No. CE170100039] and the Australian Research Council Centre of Excellence for Engineered Quantum Systems [Project No. CE170100009]. This research made use of the Rocket High Performance Computing service at Newcastle University.

\appendix

\section{Identifying the Superfluid Fraction}
\label{Appendix_superfluid}
\subsection{Decomposing the Momentum of the Wavefunction}
The momentum of the wavefunction $\Psi$ can be calculated using the relationship \cite{Pethick_and_Smith} 
\begin{equation}
    \mathbf{J}(\rr) = \frac{\hbar}{2mi} \left[ \Psi^*(\rr) \nabla \Psi(\rr) - \Psi(\rr) \nabla \Psi^*(\rr) \right].
    \label{eqn_wavefunction}
\end{equation}
Using Landau's two--fluid model we may assume that the wavefunction comprises of a superfluid component, which flows without energy loss, and a normal fluid component, which is subject to viscous effects. In this framework, the superfluid component has velocity $\vv_s$, the normal fluid component has velocity $\vv_n$, and we may write
\begin{equation}
    \mathbf{J} = \rho f_s \mathbf{v}_s + \rho f_n \vv_n, 
\end{equation}
where $f_s$ and $f_n$ are the superfluid and normal fluid fractions respectively. We now assume that the normal fluid moves with the barriers \cite{WrightSuperfluidity}, so that in the barrier frame of reference $\vv_n=0$ and $\mathbf{J}=\rho f_s \vv_s$. Since the superfluid velocity is locked to the condensate velocity \cite{Griffin,StatPhys2}, it is then relatively straight forward to calculate the average momentum of the wavefunction $\mathbf{J}$, calculate the velocity of the condensate mode, $\vv_0$, as described in Eqn.~\eqref{sf_velocity}, and extract an estimate for $f_s$.

\subsection{Using current--current correlations}
It is possible to extract the superfluid fraction of a system using the current--current correlations of the wavefunction. This result is derived in Refs.~\cite{Foster2010,Foster_hydro}, and may also be derived using the theory of hydrodynamics in a  superfluid \cite{Pit_and_String,Baym69}. Here, we summarize the approach described in these previous works to give a self-contained result.

In the limit of vanishing momentum, we write the current--current correlations of a system with volume $V$ in equilibrium at temperature $T$ as
\begin{equation}
    J_{\alpha \beta}(\kk) = \langle \left[\mathcal{F} \left( \mathbf{J} \right) \right]_\alpha \left[\mathcal{F} \left( \mathbf{J} \right) \right]_\beta^* \rangle = \left( f_s  \frac{k_\alpha k_\beta}{k^2} + f_n  \delta_{\alpha \beta}\right) \frac{k_B T V\rho}{m^2},
\end{equation}
where $\mathcal{F}\left(\mathbf{J}\right)$ indicates that the momentum is calculated using Eq.~\eqref{eqn_wavefunction} and then transformed into Fourier space  \cite{Gawryluk2019}. The current--current correlations in the system are captured by
\begin{equation}
    \chi \left( \kk \right) = \left[ \begin{matrix} J_{xx} & J_{xy} \\ J_{yx} & J_{yy} \end{matrix} \right]  = \left[ \left(f_s + f_n \right) \boldsymbol{\hat{k}} \boldsymbol{\hat{k}} + f_n \left( I - \boldsymbol{\hat{k}} \boldsymbol{\hat{k}}\right)\right] \frac{k_B T V\rho}{m^2}
\end{equation}
 where we introduce the dyad
\begin{equation}
    \boldsymbol{\hat{k}\hat{k}} = \frac{1}{k^2} \left[ \begin{matrix} k_x^2 & k_x k_y \\ k_y k_x & k_y^2 \end{matrix} \right],
\end{equation}
and $I$ is the identity. We now introduce transverse, $\chi_t(k)$, and longitudinal, $\chi_l(k)$, functions which are scalars depending only on $k$ so that 
\begin{equation}
    \chi \left( \kk \right) = \chi_l(k) \boldsymbol{\hat{k}\hat{k}} + \chi_t(k) \left( I - \boldsymbol{\hat{k}} \boldsymbol{\hat{k}}\right).
    \label{eqn:chi_decomposition}
\end{equation}
 As suggested by Eqn.~\eqref{eqn:chi_decomposition}, it is possible to identify the transverse and longitudinal parts of $\chi$ since $\chi_l(k) = \boldsymbol{\hat{k}} \cdot \chi(\kk) \cdot \boldsymbol{\hat{k}}$ and $\chi_t(k) = \boldsymbol{\hat{k_\perp}} \cdot \chi(\kk) \cdot \boldsymbol{\hat{k_\perp}}$, where $\boldsymbol{\hat{k}}$ and $\boldsymbol{\hat{k_\perp}}$ are mutually orthogonal unit vectors.

We are able to evaluate $\chi$ at all points in our system, and use the decomposition described above to find $\chi_l$ and $\chi_t$, while projecting azimuthally so that the functions depend only on $k$. Once this has been obtained, we fit each of $\ln\chi_t(k)$ and $\ln\chi_l(k)$ to a quadratic function. As our simulations are computed on a square grid, the density of points increases with $k$; to account for this in our curve fitting procedure, we set the uncertainty to be $k^{1/2}$ (equivalent to a $1/k$ weighting in the fit). These procedures follow those of Ref.~\cite{Foster2010}.

Finding the normal fluid density corresponds to taking the limit as $k\to 0$ of the transverse component of $\chi$, while the same limit of the longitudinal component of $\chi$ gives the sum of the superfluid and normal fluid densities. Taking the limit as $k\to 0$ of the quadratic function for the parameters found from our curve fitting procedure allows us to calculate the normal fluid fraction as
\begin{equation}
    \frac{f_n}{f_s + f_n} = \frac{\lim_{k\to 0} \chi_t(k)}{\lim_{k\to 0} \chi_l(k)}.
\end{equation}
 This allows us to relate the superfluid and normal fluid fractions to correlations from our simulations, in a similar manner to the condensate and non-condensate fractions which are determined using $G^{1B}$.

\section{The Ring--Plaquette Method}
\label{appendix_ring_plaquette}
In this section we describe the method used to detect the winding number, $\mathcal{W}_k$, about a given barrier. 

\begin{figure}
    \centering
    \includegraphics{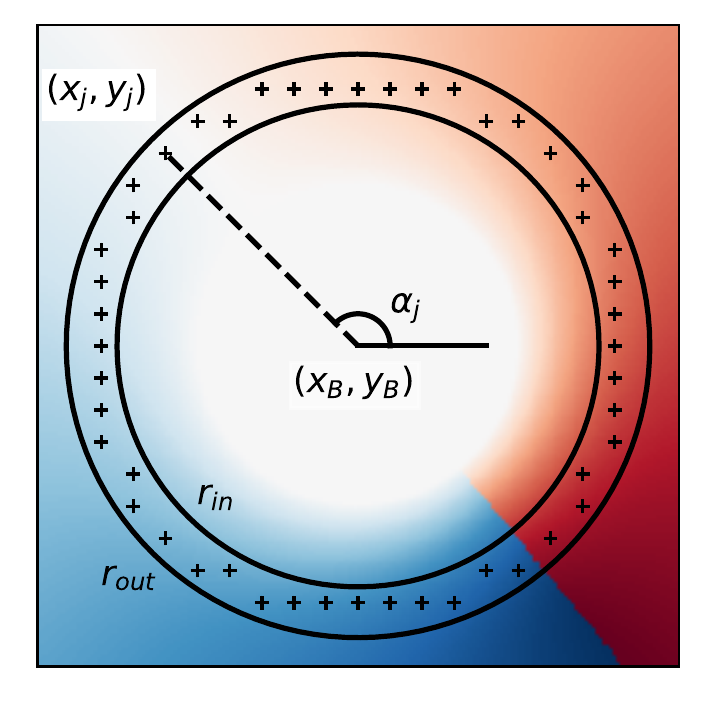}
    \caption{A schematic of the numerical method used to compute the winding number in each barrier. The contour integral in Eqn.~\eqref{eqn_phase_integral} is performed by evaluating the phase at the grid points (indicated with pluses) located within the annulus with inner radius $r_{in}$ and outer radius $r_{out}$,  shown. The angle $\alpha_j$ at each grid point can be computed by $\alpha_j = \text{arctan2}\left(y_B-y_j,x_B-x_j\right)$. The colour plot represents a density weighted plot of the phase, $\theta|\Psi|^2 $, where $\Psi= |\Psi|e^{i\theta}$; the white central region represents the area within the barrier where the density of the fluid vanishes. In this case, the barrier supports a winding number $\mathcal{W}_k=1$.}
    \label{fig:cartoon_ring}
\end{figure}

Using the Madelung transformation, we write the wavefunction as $\Psi = |\Psi|\exp\left(i\theta\right),$ where $|\Psi|^2$ is the particle density, and $\theta$ is proportional to the velocity potential.  The circulation of a quantum fluid is quantized, so that around any closed contour enclosing barrier $k$, and no other vortices, the change in the phase, $\Delta \theta$ is given as
\begin{equation}
    \Delta \theta = \oint_\mathcal{C} \nabla \theta \cdot d\rr = 2\pi \mathcal{W}_k
    \label{eqn_phase_integral}
\end{equation}
for some integer $\mathcal{W}_k$ which we shall refer to as the winding number. 

In our simulations, the wavefunction $\Psi$ is computed at discrete grid points and so we calculate the line integral in Eqn.~\eqref{eqn_phase_integral} numerically. For a barrier with centre $\left(x_B,y_B\right)$ and effective width $a$, we create an annulus which has inner radius $r_{in}$ and outer radius $r_{out}$. A sketch of this set-up is given in Fig.~\ref{fig:cartoon_ring}. The inner and outer radii are chosen so that the computational grid points contained within the annulus are outside the zero density region of the barrier, but do not overlap with the annuli enclosing other barriers. Once the grid points contained within the annulus have been identified, they are sorted in order of increasing angle $\alpha_j$ and the phase of the wave function is evaluated at each point. We then calculate the unwrapped phase difference between neighbouring points,
\begin{equation}
    \Delta \theta_{j,j+1} = \theta \, \big|_{\alpha_j} - \theta \, \big|_{\alpha_{j+1}} .
\end{equation}
It is necessary to unwrap the phase in this way to ensure that the phase is continuous between neighbouring points \cite{Foster2010}, however working on a discrete grid this continuity is poorly defined as there may be jumps in the phase of $2\pi$; to correct for this we add multiples of $2\pi$ so that $|\Delta \theta_{j,j+1}|<\pi$ . The winding number is then computed as 
\begin{equation}
    \mathcal{W}_k = \frac{1}{2\pi}\sum_{j} \Delta \theta_{j,j+1}.
\end{equation}

\bibliography{disordered_flow_references}

\begin{thebibliography}{93}
\bibitem [{FFT()}]{FFT_footnote}%
  \BibitemOpen
  \href@noop {} {\enquote {\bibinfo {title} {{Note that any aliasing problems
  which may have arisen when computing the first derivative term in
  Eqn.~\eqref{moving_gpe} using Fast Fourier Transforms are corrected by the
  cut-off implemented by the projection operator \cite{fft_info}.}}}\
  }\BibitemShut {NoStop}%
\end{thebibliography}
\end{document}